\documentclass[preprint]{elsarticle}
\journal{Applied energy}

\usepackage{amssymb}
\usepackage{amsthm}
\usepackage{amsmath}
\usepackage[shortlabels]{enumitem}
\usepackage{caption}
\usepackage{subcaption}
\usepackage{float}
\usepackage[dvipsnames]{xcolor}

\begin{document}

\begin{frontmatter}
\title{One-class anomaly detection through color-to-thermal AI for building envelope inspection}

\author[DI]{Polina Kurtser}
\author[IHBI]{Kailun Feng}
\author[AF]{Thomas Olofsson}
\author[F]{Aitor De Andres\corref{cor1}}
\ead{aitor.de.andres@umu.se}

\cortext[cor1]{Corresponding author}

\affiliation[DI]{organization={Department of Diagnostics and Intervention, Radiation Physics, Umeå University},
            addressline={5A, H62, Norrlands Universitetssjukhus}, 
            city={Umeå},
            postcode={901 85}, 
            country={Sweden}}

\affiliation[IHBI]{organization={Intelligent Human-Buildings Interaction Lab, Department of Applied Physics and Electronics, Umeå University},
addressline={X, Håken Gullessons väg 20}, 
city={Umeå},
postcode={901 87}, 
country={Sweden}}

\affiliation[AF]{organization={Department of Applied Physics and Electronics, Umeå University},
addressline={X, Håken Gullessons väg 20}, 
city={Umeå},
postcode={901 87}, 
country={Sweden}}

\affiliation[F]{organization={Department of Physics, Umeå University},
addressline={Linnaeus väg 24}, 
city={Umeå},
postcode={901 87}, 
country={Sweden}}

\begin{abstract}

We present a label-free method for detecting anomalies during thermographic inspection of building envelopes. It is based on the AI-driven prediction of thermal distributions from color images. Effectively the method performs as a one-class classifier of the thermal image regions with high mismatch between the predicted and actual thermal distributions. The algorithm can learn to identify certain features as normal or anomalous by selecting the target sample used for training. We demonstrated this principle by training the algorithm with data collected at different outdoors temperature, which lead to the detection of thermal bridges. The method can be implemented to assist human professionals during routine building inspections or combined with mobile platforms for automating examination of large areas.

\end{abstract}

\end{frontmatter}

\section{Introduction}

With various sustainability targets quickly approaching, an important effort is underway to improve the energy efficiency of buildings, the single largest energy consumer worldwide \cite{IEA2022, EU2023}. Instead of focusing on improving design standards for new buildings, which constitute only about 1\% of the total stock per year \cite{EMF2023}, the attention has steadily shifted to selectively refurbishing existing buildings and their most inefficient components \cite{d2018technical}. However, characterizing the energy performance of each component and locating anomalies is not an easy task: most buildings were constructed before proper efficiency standards were enforced and detailed energy assessments are often missing \cite{feng2022energy}. On top of that, energy inefficiencies may also arise in newer buildings due to factors such as lack of maintenance, improper installation, and natural degradation. In this context, energy authorities and retrofitting companies need available, effective, and fast methods to evaluate buildings and identify inefficient or faulty components.

Passive infrared thermography (PIRT) is a powerful tool routinely used in the context of civil engineering and building inspection \cite{kylili2014infrared}. It relies on imaging the thermal radiation emitted by a target scene onto a two-dimensional infrared (IR) detector and can, under certain assumptions, retrieve the spatially-resolved target temperature. It has been proven effective in locating energy efficiency anomalies in buildings such as thermal bridges, air leaks, structural faults, and mould \cite{kylili2014infrared, martin2022infrared, abdelhafiz2022innovative}, as well as useful to measure the environmental sol-air temperature \cite{ohlsson2018step, ohlsson2019sol}. However, PIRT images interpretation requires significant domain knowledge and is prone to measurement artifacts arising from a complex interplay of factors such as diffuse and specular reflections, angle of emission, wind speed, sky temperature, and changes in emissivity \cite{minkina2009infrared, lehmann2013effects, ohlsson2014quantitative}. As a result, despite recent algorithmic advances, the execution of building inspection with PIRT remains labor-intensive and time-consuming, with skilled professionals required to interpret the data and discard artifacts \cite{fox2016building}. While traditional algorithms are commonly employed to process and enhance thermographic images, their success when it comes to automating their interpretation has so far been limited, especially in situations of high variability \cite{garrido2020thermographic}. However, the rapid development of artificial intelligence (AI) during the past decade has revolutionized machine vision, offering vast opportunities that the field of thermography for building inspection has so far exploited to a very limited degree \cite{shariq2020revolutionising, panella2021brief}.

The existing literature devoted to the intersection between AI and thermographic analysis of building envelopes is rather limited \cite{sabato2023non, gertsvolf2021aerial}. Current state-of-the-art focuses primarily on supervised learning approaches - methods that rely on the meticulous annotation of RGB and thermal images, attempting to encompass the variability using the acquisition of large datasets. Garrido et al.\cite{garrido2023introduction} presented a method for thermal bridges and water-related problems detection by employing a deep learning model (Mask R-CNN). To improve the network detection performance, they introduce a pre-processing step to the contrast between anomalous and non-anomalous areas of the building envelope. Yang et al. studied different multimodal RGB-thermal fusion techniques  \cite{yang2023comparison}. They concluded that using a UNet architecture to perform encoding of a common latent space of the RGB and thermal images is beneficial for detection of thermal issues such as detaching and missing tiles in building facades.
Alexakis et al. \cite{alexakis2024novel} showed how these deep network models can be utilized for early identification of rising dampness at historical monuments. In addition, literature in laboratory conditions includes the work of Royuela et al., \cite{royuela2019air} who developed a method for the assessment of air infiltration using convolutional and fully connected neural network models. Similarly, Fang et al.,\cite{fang2023automatic} used supervised, labeled based deep learning architectures for detection of several anomalies.

 It is difficult to pin-point why the literature devoted to applying AI methods to thermal building envelope inspection is limited, especially in comparison to other automation applications heavily relying on computer vision, such as autonomous navigation, robotic perception, remote and local sensing, and precision medical imaging. The low amount of publicly available datasets acquired in uncontrolled conditions, the requirement of extensive labeling, as well as the knowledge gap between building experts and AI professionals have been discussed in the literature as possible causes \cite{sabato2023non, shariq2020revolutionising, gertsvolf2021aerial}. Furthermore, to the best of the author's knowledge machine learning and AI tools have not yet been applied in commercial building inspection, which suggests the mentioned proof of concept models do not generalize well to the variability of realistic scenarios. This lack suggests that commonly applied, data-driven, supervised learning methods should potentially be replaced with alternative methods of anomaly detection, less driven by expert knowledge. Thus, a knowledge gap exist regarding how new or existing AI methods can be applied to the inspection of building envelopes.

To bridge that gap, we build on an interesting approach reviewed by Pang et al. for the deep representation of normality \cite{pang2021deep}. Specifically, the authors review several methods for learning feature representations of normality using deep autoencoders (AE) and generative (GAN) architectures.  This is a relatively recent development of the self-representation theory – if one builds a model trained to represent normal data it will fail to represent abnormalities. A deep auto-encoder-based anomaly detection method uses an encoder (set of densely connected layers with stepwise decreasing dimensionality) to first learn to map data samples into a latent space (the last layer of the encoder with the smallest dimension) and then reconstruct them to the original images by a decoder (another set of dense layers stepwise increasing in dimensionality). Hence, auto-encoder trained on normal samples would reconstruct normal images well but would fail to reconstruct abnormal images. The reconstruction error (an error map between the original and reconstructed images) is used as a metric for anomaly detection. These methods show promising performances on benchmark datasets \cite{yan2021learning}, while maintaining a main strength - eliminating the need to label anomalies.

In this paper, we present a development on the deep representation of normality approach by applying it to RGB-thermal image pairs. The technique aims to overcome the stated challenges, especially for the generalization problem and applicability to building inspection in uncontrolled environments, data acquisition, and data labeling. We demonstrate it specifically for the location of thermal bridges in building facades, but it can similarly be utilized for the location of other anomalies. The technique can be used as it is for assisting human professionals during manual inspections, or in conjunction with mobile platforms and other classification schemes for the automated inspection of large urban areas. 

The article proceeds by briefly introducing the necessary AI frameworks and tools. Further, it introduces and formalizes the algorithm structure. After that, it goes into the detail of image acquisition, preprocessing, and network training. Finally, the results are presented and discussed.

\section{Introduction to relevant AI frameworks}

    \subsection{Generative-adversarial networks}
    Generative deep learning networks are networks capable of generating new samples closely resembling the distribution of the input data. While a number of generative paradigms exists in the literature, in this paper we focus on  Generative-adversarial networks (GAN) - generative models in which two networks, the generator and the discriminator, compete to outperform each other \cite{goodfellow2014generative, gui2021review, dubey2023transformer}. Given a training data set T, the goal of the generator network is to learn how to model T and to generate new synthetic data S with the same statistical distribution as T. Meanwhile, the goal of the discriminator network is to identify if a test data sample belongs to either T or to S. During the training process, both networks evolve in parallel as the generator learns how to fool the discriminator, and the discriminator learns how to not be fooled by the generator. Nowadays, GAN algorithms are commonly used in applications requiring the generation of highly realistic data.

    \subsection{Image-to-image translation networks}
    Image-to-image translation or style networks build upon the generative aspects of GAN to perform domain transfer by generating an image in one domain to an image in another. The idea of training a generative architecture on a non-identical input-output image pair (thus not auto encoding but rather image-to-image translation), has been explored in a number of network architectures. For example, the generator of CycleGAN \cite{zhu2017unpaired} can be realized through a ResNet or U-net architectures. Both consist of encoder and decoder blocks realized in convolutional layers and skip connections. While the ResNet based generator also introduces a multi-modal translation part between the encoding and decoding consisting of residual blocks. Some generative architectures such as pix2pix \cite{isola2017image}, learns a mapping from input images to output images while the mentioned above CycleGan architecture, an extension of the pix2pix lifts the need to have paired multi-modal data samples. Extending even further on this idea recent Transformer-based GANs utilize a self-attention mechanism, originally proposed for a machine translation task in the generator part, where the encoder and decoder networks are built using transformers \cite{dubey2023transformer}. 

    \subsection{Color-to-thermal networks}

    Color-to-thermal networks are a specific application of image-to-image networks whose purpose is to translate an input color image (typically 3 channels, RGB) to their corresponding thermal distribution (single channel). Kniaz et al. successfully demonstrated this approach using a dedicated network, thermalGAN, to predict the thermal distribution of different objects \cite{kniaz2018thermalgan}. Later, Mizginov tested and compared the performance of different GAN architectures when used as color-to-thermal \cite{mizginov2021method}. Both studies included buildings as a possible generation category and, thus, hold great promise for the study of building facades.

   \subsection{One-class classification}
    One-class classification (OCC) is an algorithm classification paradigm that aims to identify test objects as either positive (a.k.a. target) or negative (a.k.a. outlier) \cite{Khan_Madden_2014}. Unlike traditional multi-class classification algorithms, which typically require a collection of statistically significant data for each possible category, most OCC algorithms are trained using only positive data, learning to model it and to recognize anything outside the trend as an outlier. This paradigm is useful when the negative data is either not available, not enough to be statistically significant, or not well defined. 
    In the context of the paper, we aim to perform the detection of heat leaks in RGB-thermal pairs using one-class classification, i.e. our model training is reliant on anomaly-free RGB-thermal pairs for the identification of the underlying normal distribution. In that sense, the training is loosely supervised where image pairs are required for the training of the color-to-thermal network, and labeling of the single class of anomaly-free thermal images is required. But annotation of anomaly instances, a major bottleneck in the implementation of these algorithms in the thermal inspection domain, is not required.

\section{Algorithm structure}

The proposed method is inspired by the intuition-based reasoning of human inspectors, who find anomalies based on experience-acquired expectations.

\begin{figure}[h]
    \centering
    \includegraphics[width = 0.8\textwidth]{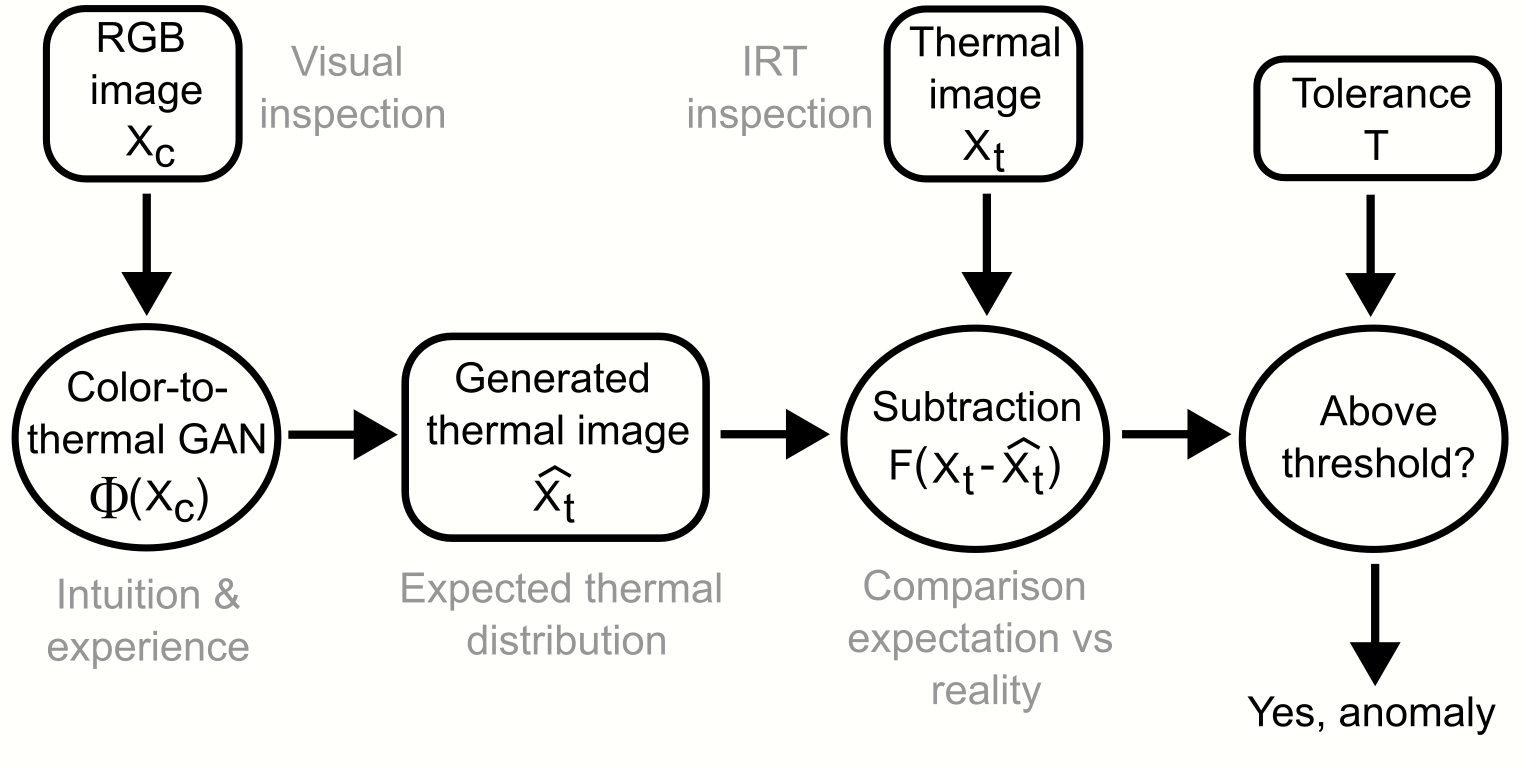}
    \caption{Algorithm workflow in black, with the corresponding human-driven steps in gray. After training A color-to-thermal GAN trained on data with the desired properties first produces the expected thermal distribution of a building envelope image (RGB). Then, the algorithm compares the expected and actual thermal distributions. Strong mismatches between both are labelled as anomalies.}
    \label{fig:algorithm_sketch}
\end{figure}

The algorithm (flowchart in Figure \ref{fig:algorithm_sketch})  utilizes pairs of aligned RGB ($x_c$) and thermal images ($x_t$) of size $MXN$ to perform the detection of heat leak instances. At first stage a mapping function $\Phi$ is learned to estimate a thermal image $\widehat{x_t}$ such as:
\begin{equation}
    \widehat{x_t} = \Phi(x_c)
\end{equation}

The mapping function is learned using a color-to-thermal translation network for the generation of the estimated thermal image. Then, an anomaly metric map ($E$) is calculated as a function of the difference between the expected thermal realization  and the thermal image as sensed by the thermal camera,

\begin{equation}
E = F(x_t-\widehat{x_t}),
\label{eq:difference}
\end{equation}

where $F$ is an identity function in the presented application.

To identify anomaly regions $A$ and subsequently detection of anomalous objects the anomaly metric $E$ is thresholded with an input tolerance $T$ to perform segmentation of the image areas with deviating anomaly metric  - i.e.,  unexpected thermal measures, which constitute as anomalies.

\begin{equation}
A_{i,j} = \left\{\begin{matrix}
0 & E_{i,j}\leqslant T   \\ 
1 & E_{i,j}>T  & 
\end{matrix}\right.\forall i\in [1..N], j\in [1..M]
\end{equation}

To estimate the mapping function $\Phi$ and generate an expected thermal realization $\widehat{x_t}$ from a corresponding RGB image $x_c$, we train a pix2pix deep learning architecture \cite{isola2017image}. The pix2pix architecture utilizes the supervised paired relations between the aligned RGB and thermal images and has been tested in the literature for the color-to-thermal image mapping problem, albeit with mixed results \cite{mizginov2021method}. Nevertheless, it has also shown to be highly successful in image-to-image translation for the generation of realistic RGB facades from categorical labels using relatively small data sets (606 images) \cite{tylecek2012cmp}. 

It is important to highlight that finding a mapping function $\Phi$ by training a pix2pix network on exclusively positive, anomaly-free data samples, is expected to encompass the relations between RGB and thermal data in anomaly free cases. Therefore, a careful selection of training data is critical for shaping the algorithm to identify the anomalies of interest in a given application. To illustrate this principle, we used data corresponding to three different environmental conditions and demonstrated how this can shape the model to identify thermal bridges in building facades.

\section{Implementation}
\subsection{Data acquisition}
Pairs of visual and thermal images were acquired using a visible (RGB) and a thermal camera mounted together on the top of a ground vehicle (a car). The visible camera was a thermally-stabilized, multi-lens, and aberration-corrected stereo camera (Stereolabs, ZED2), with a field of view of 110$^o$ horizontal x 70$^o$ vertical and a sensor resolution of 2x1920x1080 px. The stereo functionality (depth-sensing) was chosen for future reference but not used in this study. The thermal camera was a single-lens, uncooled, and mechanically shuttered microbolometer (Seek thermal, Mosaic Core), with a field of view of 56$^o$ horizontal x 42$^o$ vertical and a sensor resolution of 320x240 px. Due to its single lens configuration, the thermal camera suffered significant imaging distortion, which was dealt with during the preprocessing phase. The acquisition took place place by driving more or less randomly in an undisclosed neighbourhood in Sweden, which was chosen for its relatively varied architecture and construction period of their buildings. Several acquisition rounds were performed at three different outdoors environmental conditions and the data split in 3 corresponding sets, as detailed in table \ref{tab:acquisition_sets}. In each set, the available images were randomly assigned for either training the neural network or for evaluation purposes. In all cases, the wind speed was low ($<4 m/s$) and there was diffuse (i.e., not direct) sun light exposure during and for at least 2 hours before the acquisition. The acquisition of the two winter data sets was performed during the day hours, with a partially cloudy ski and high coverage of snow and ice on the street ground. The acquisition of the summer set was performed at sunset time and with clear sky.

\begin{table}[h]
    \centering
    \caption{Data sets and acquisition conditions}
    \begin{tabular}{c|c|c|c}
        Set name & Winter4 & Winter8 & Summer \\
        Outdoors temperature ($^o C$) & -4 & -8 & 17 \\
        Number training images & 948 & 865 & 184 \\
        Number evaluation images & 50 & 50 & 9 \\
    \end{tabular}
    \label{tab:acquisition_sets}
\end{table}

\subsection{Preprocessing}

After acquisition, the thermal and visual images were preprocessed in four steps. First, the imaging distortion was removed from the thermal data using a second order radial (barrel) model. The distortion parameters were first guessed using using the method described in \cite{zhang2000flexible} with a heated chessboard pattern and then manually fine-tuned to optimize their alignment (i.e., their overlap with the RGB images). Second, the visual images were cropped to the same field of view of the thermal camera and both were interpolated to a spatial resolution of 512x512 px. Third, the thermal data (absolute temperature, float precision) was encoded by making it relative to the outdoors temperature in each case, blanking it outside the interval [-5, 10] $^oC$, rounding it to reduce its resolution to $0.5$ $^oC$, and shifting it to the positive integer space. Notice that after this process, each pixel in the thermal images can only take integer values between 0 and 30. Fourth and last, both visual and encoded thermal data were saved as 8-bit (png) images.

\subsection{Network training}

Supervised training of the network model was performed using the pix2pix Matlab implementation by \cite{Pinkney2024}. We trained three neural models, one for each data set, using the parameters detailed in table \ref{tab:training_parameters}, with the rest being the standard of the implementation. Data augmentation was implemented by horizontal mirroring. While the networks Winter4Net and Winter8Net were trained from the scratch using the respective data sets, the same approach could not be used for the Summer network due to the low amount of available images. In this case, we used the trained network Winter4Net as a starting point for the SummerNet, performing only a rather soft retraining with the Summer data set.

\begin{table}[h]
    \centering
    \caption{Training parameters}
    \begin{tabular}{c|c|c|c}
        Network name & Winter4Net & Winter8Net & SummerNet \\
        Generator learning rate & $10^{-3}$ & $10^{-3}$ & $2\cdot10^{-4}$ \\
        Discriminator learning rate & $10^{-3}$ & $10^{-3}$ & $2\cdot10^{-4}$ \\
        Number epochs & 150 & 150 & 30 \\
        Mini-batch size & 3 & 3 & 3 \\
    \end{tabular}
    \label{tab:training_parameters}
\end{table}

\section{Results and discussion}

\begin{figure}[h!]
    \centering
    \begin{subfigure}[b]{0.575\textwidth}
        \centering
        \includegraphics[width=\textwidth]{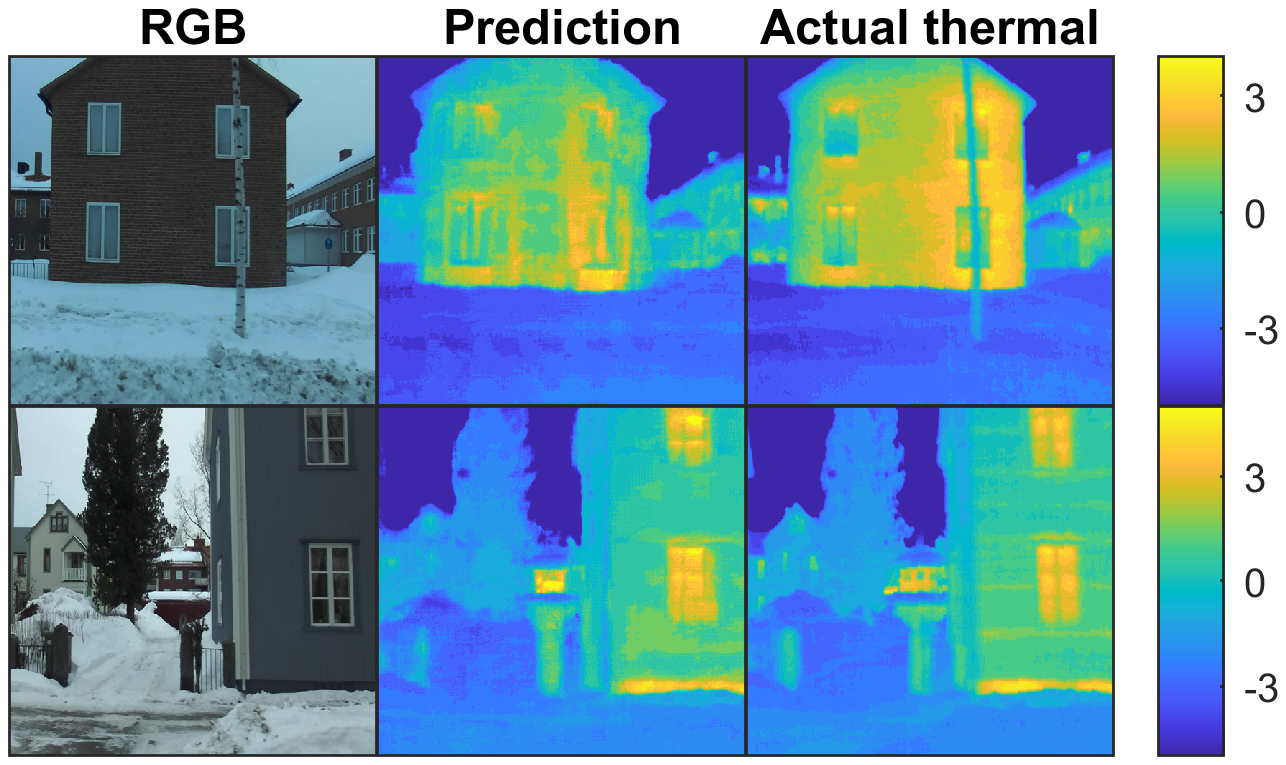}
        \caption{}
        \label{fig:prediction_Net4_on_Net4}
    \end{subfigure}
    \hfill
    \begin{subfigure}[b]{0.375\textwidth}
        \centering
        \includegraphics[width=\textwidth]{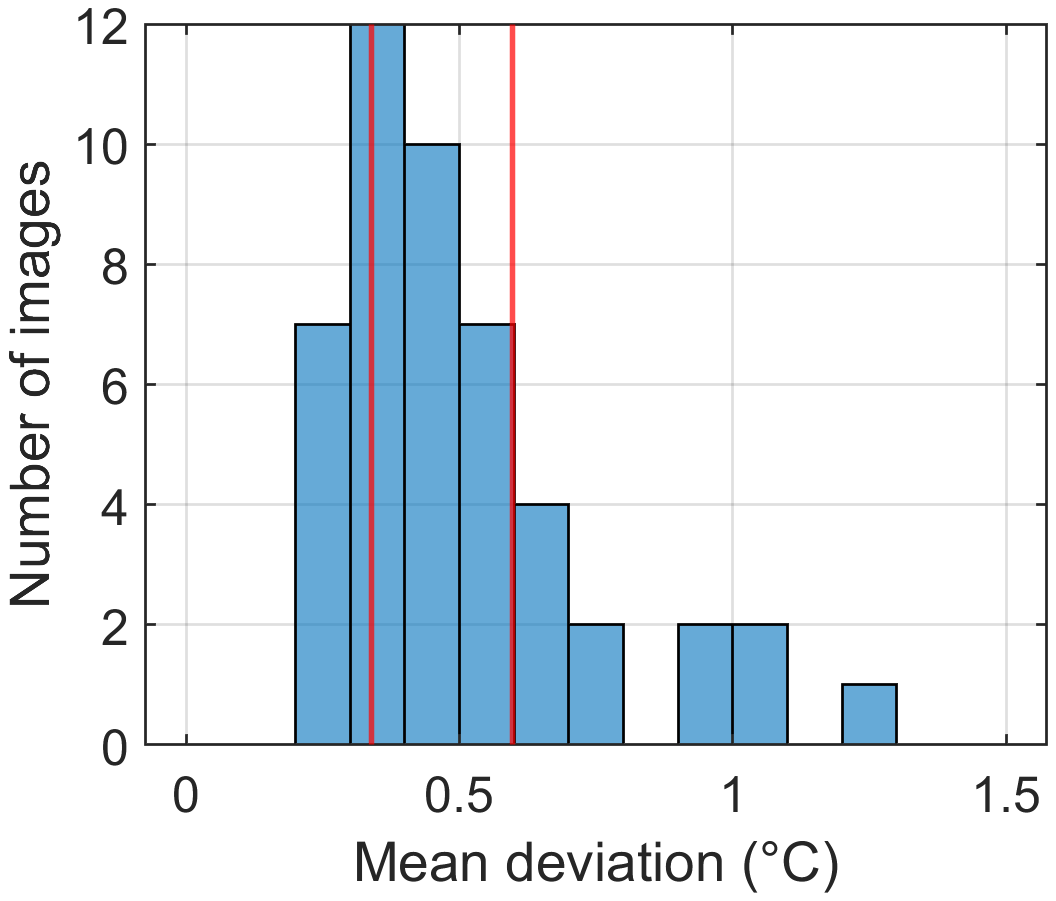}
        \caption{}
        \label{fig:stats_Net4_on_Net4}
    \end{subfigure}
    \hfill
    \begin{subfigure}[b]{0.575\textwidth}
        \centering
        \includegraphics[width=\textwidth]{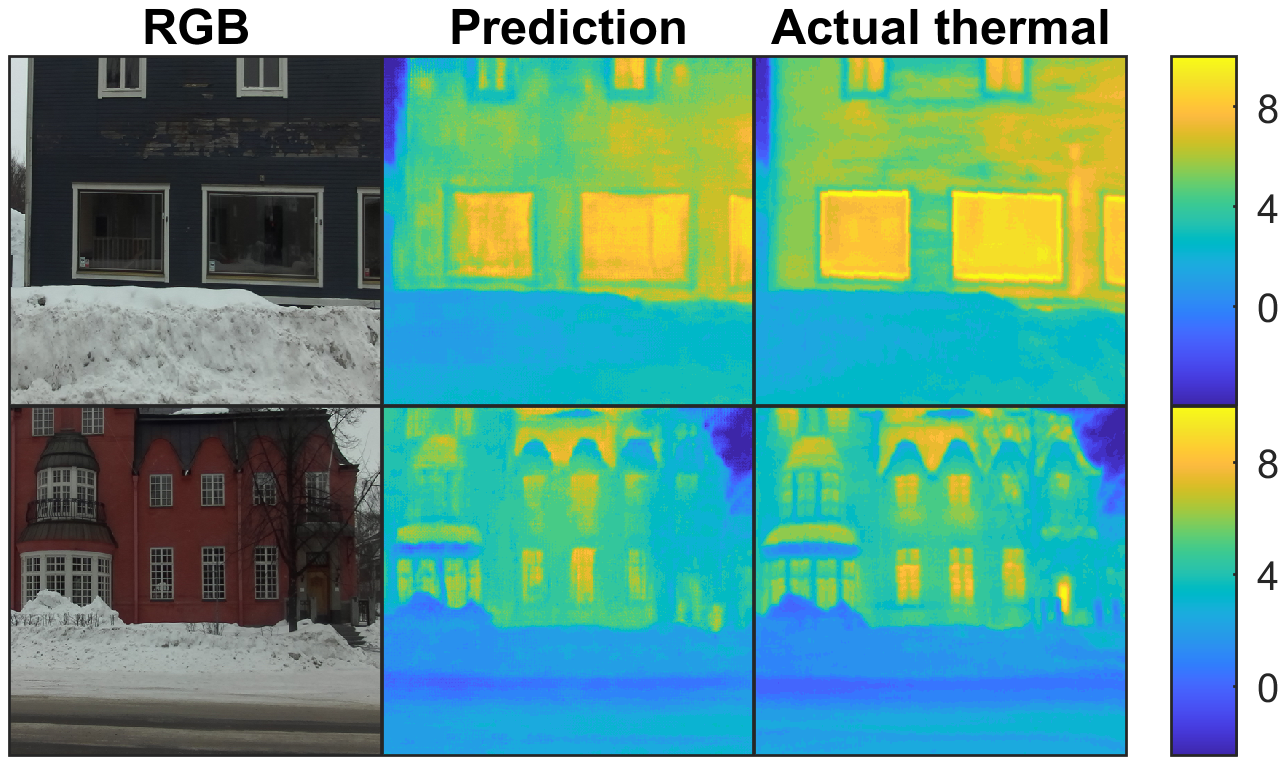}
        \caption{}
        \label{fig:prediction_Net8_on_Net8}
    \end{subfigure}
    \hfill
    \begin{subfigure}[b]{0.375\textwidth}
        \centering
        \includegraphics[width=\textwidth]{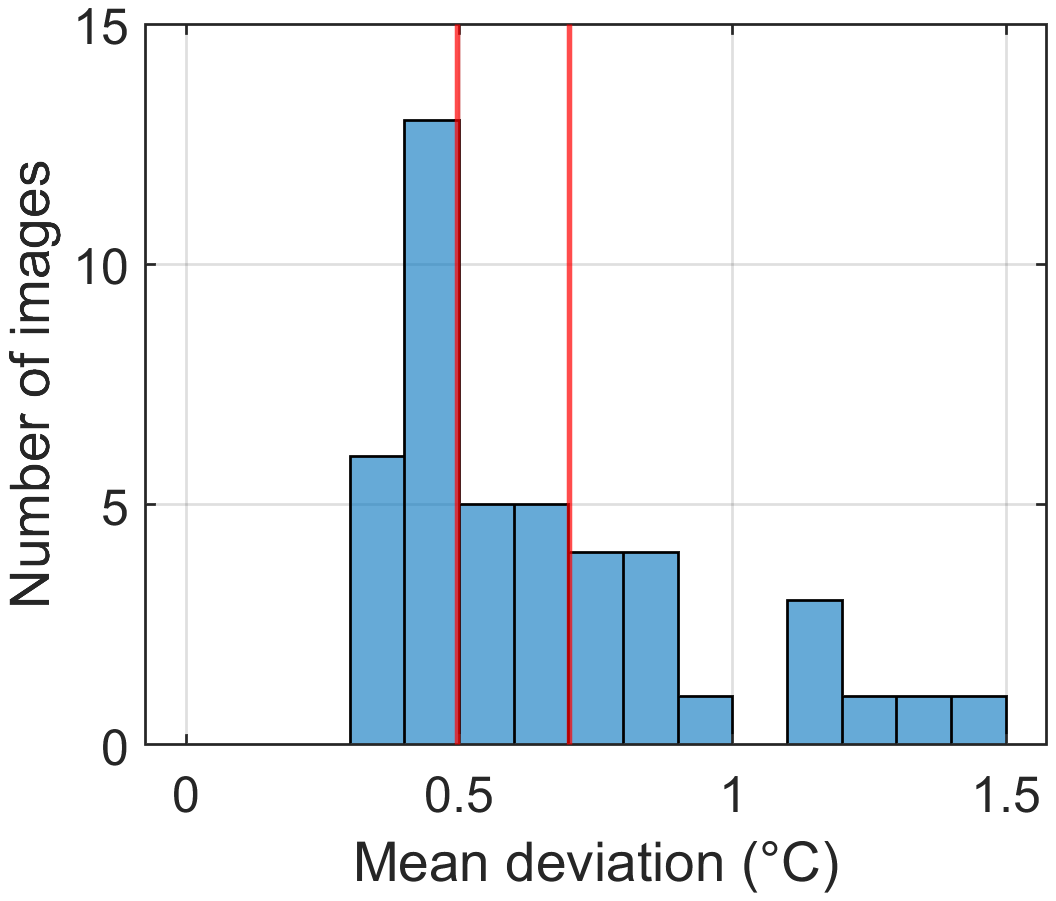}
        \caption{}
        \label{fig:stats_Net8_on_Net8}
    \end{subfigure}
    \hfill
    \begin{subfigure}[b]{0.575\textwidth}
        \centering
        \includegraphics[width=\textwidth]{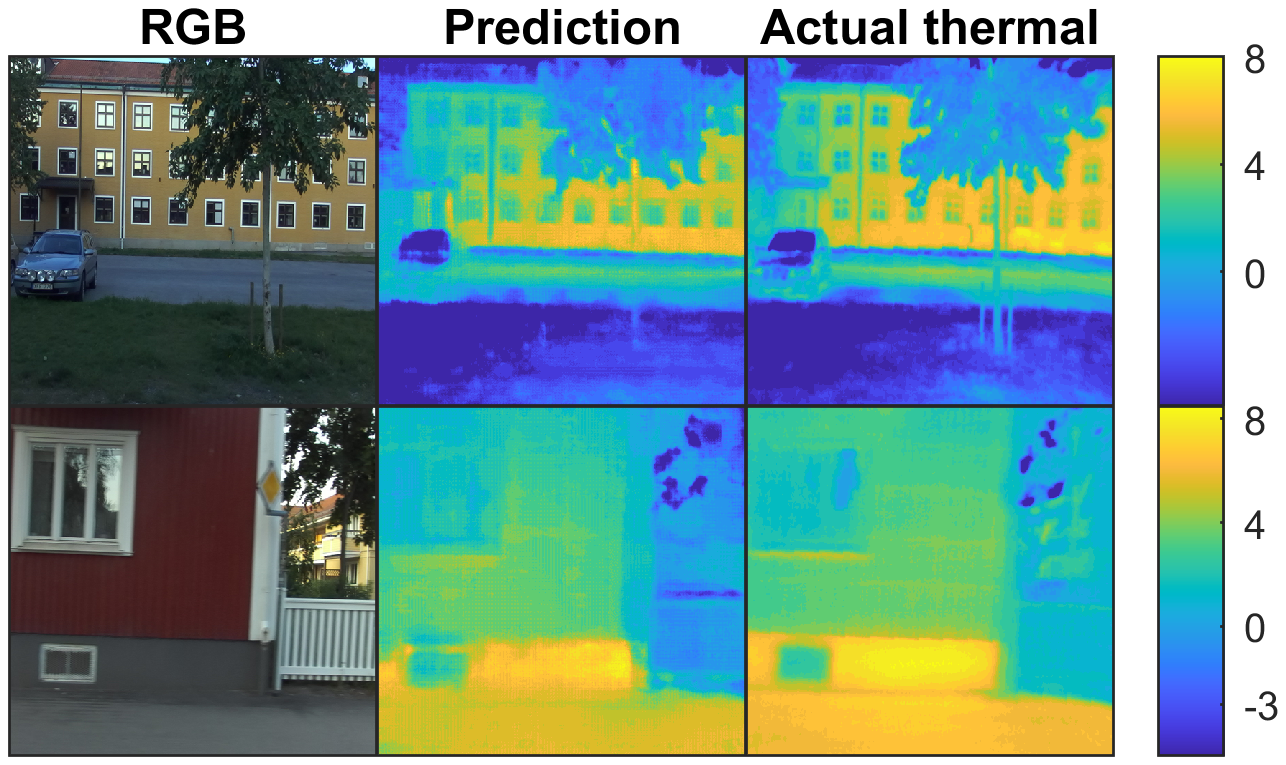}
        \caption{}
        \label{fig:prediction_sum_on_sum}
    \end{subfigure}
        \hfill
    \begin{subfigure}[b]{0.375\textwidth}
        \centering
        \includegraphics[width=\textwidth]{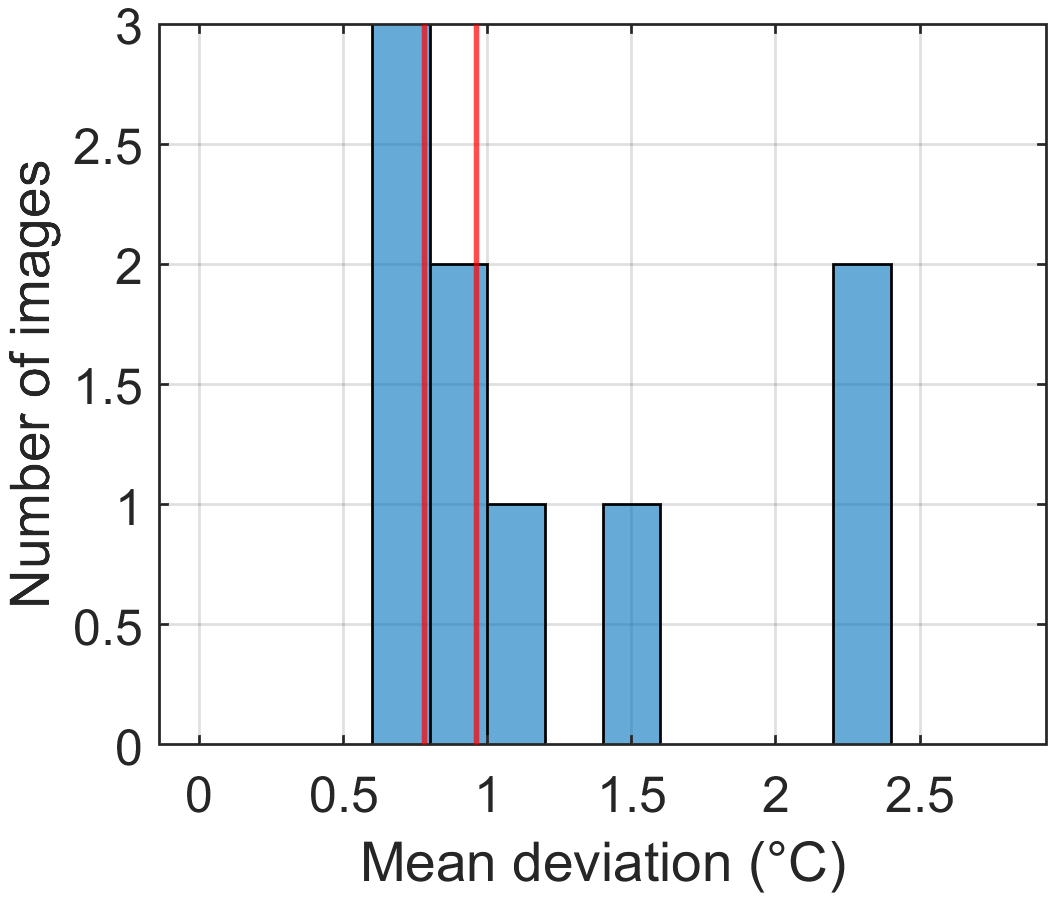}
        \caption{}
        \label{fig:stats_sum_on_sum}
    \end{subfigure}
    \caption{Performance evaluation of the trained networks (a,b) winter4Net, (c,d) winter8Net, and (e,f) SummerNet  when making predictions on their own test data sets. The histogram plots show the average pixel deviation for all the images in the corresponding test sets, with the red vertical lines indicating the deviation for the displayed images. Scale bars in $^o$C over the outdoors temperature.}
    \label{fig:results_own_test_data}
\end{figure}

We started by analyzing the performance of the network model to learn the correlation between RGB images and their thermal distributions for given environmental conditions. Figure \ref{fig:results_own_test_data} shows the comparison between the actual and predicted thermal images produced by the networks Winter4Net (\ref{fig:prediction_Net4_on_Net4}), Winter8Net (\ref{fig:prediction_Net8_on_Net8}), and SummerNet (\ref{fig:prediction_sum_on_sum}) when evaluating the RGB images of their corresponding test sets. Notice that the test data was not included in the training in any case. The accuracy in the prediction performs well in many cases  and exceeds the results presented by \cite{mizginov2021method}. We speculate that this improvement could be due to a more optimal preprocessing applied in our case before training, which we observed had great influence in the network performance. More specifically, the higher spatial resolution, narrow dynamic range, and reduced temperature precision were critical to achieve good results. 

Qualitatively, the model seems to identify the windows of older looking buildings as warm during the winter, due to a heat-flux from the indoors to the outdoors, but cold during the summer, when there is little to no indoor-outdoor heat-flux. It also learned to identify most basements as mildly warm; Swedish residential buildings tend to have common laundry equipment in the basement, which produces significant amount of heat all year round.

\begin{figure}[h!]
    \centering
    \begin{subfigure}[b]{0.6\textwidth}
        \centering
        \includegraphics[width=\textwidth]{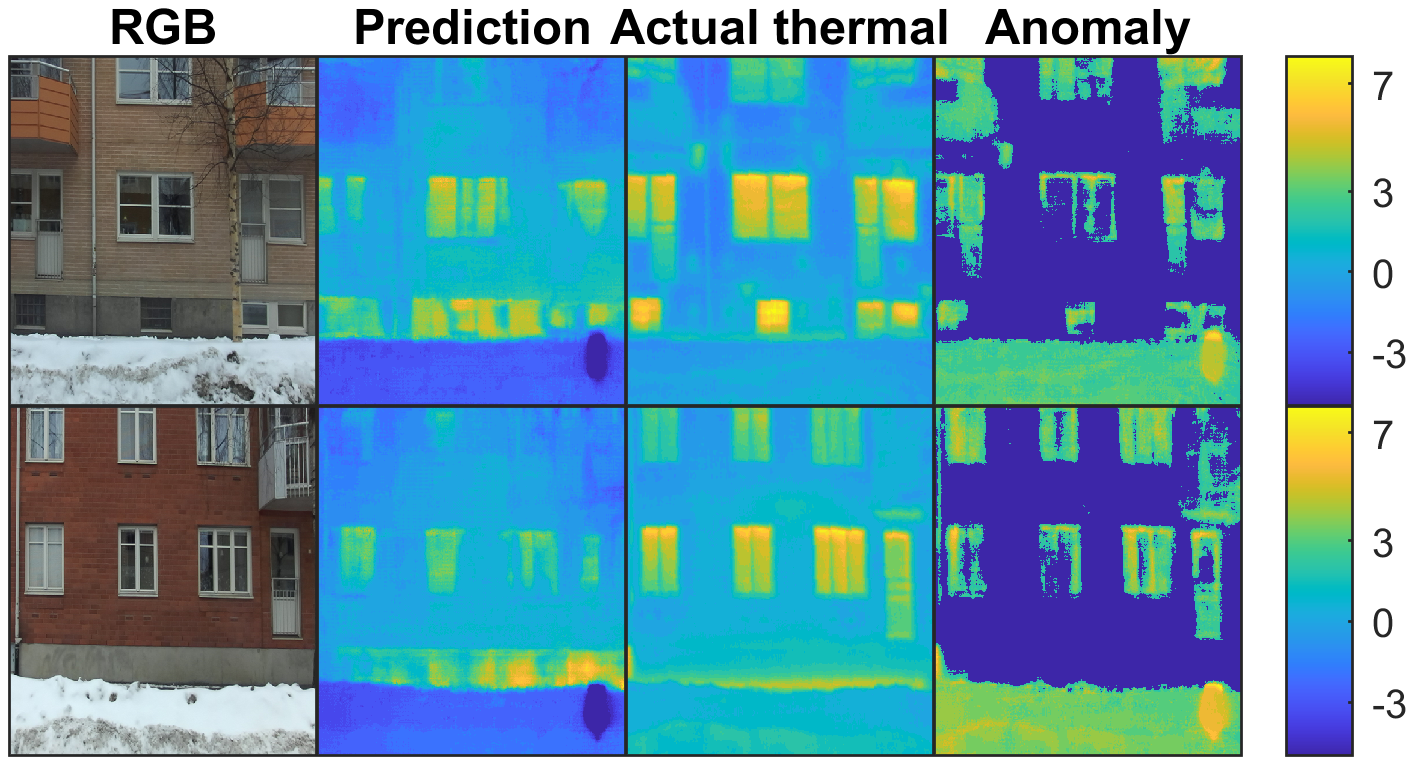}
        \caption{}
        \label{fig:prediction_Wint4_on_Wint8}
    \end{subfigure}
    \hfill
    \begin{subfigure}[b]{0.35\textwidth}
        \centering
        \includegraphics[width=\textwidth]{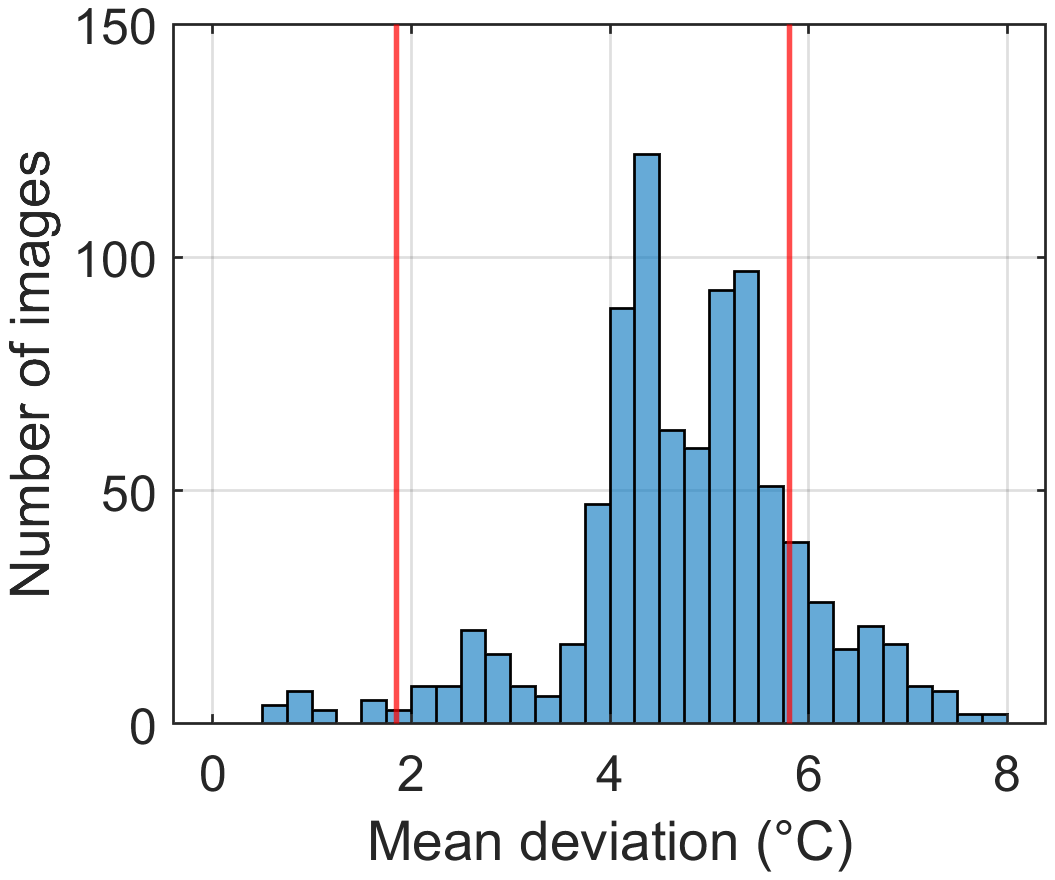}
        \caption{}
        \label{fig:stats_Wint4_on_Wint8}
    \end{subfigure}
    \hfill
    \begin{subfigure}[b]{0.6\textwidth}
        \centering
        \includegraphics[width=\textwidth]{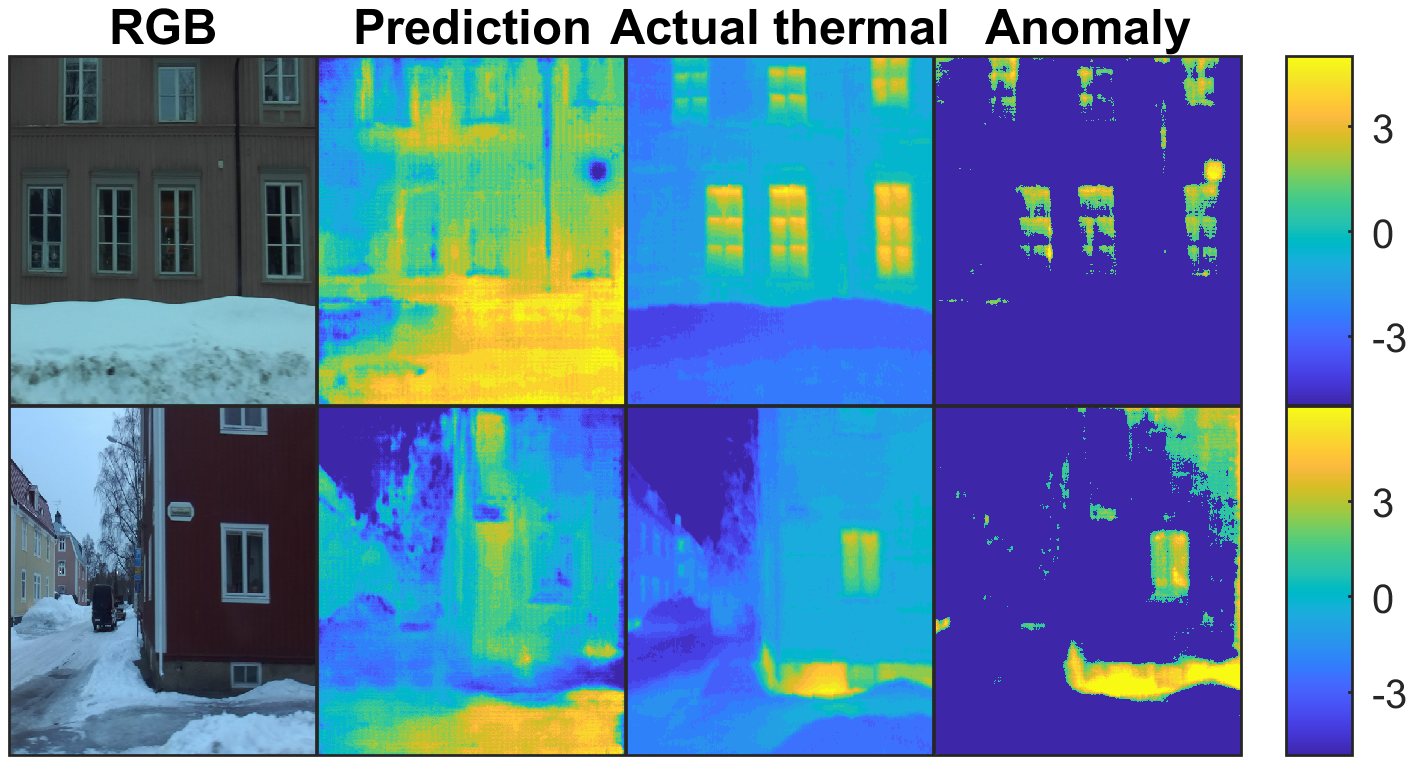}
        \caption{}
        \label{fig:prediction_Sum_on_Wint4}
    \end{subfigure}
    \hfill
    \begin{subfigure}[b]{0.35\textwidth}
        \centering
        \includegraphics[width=\textwidth]{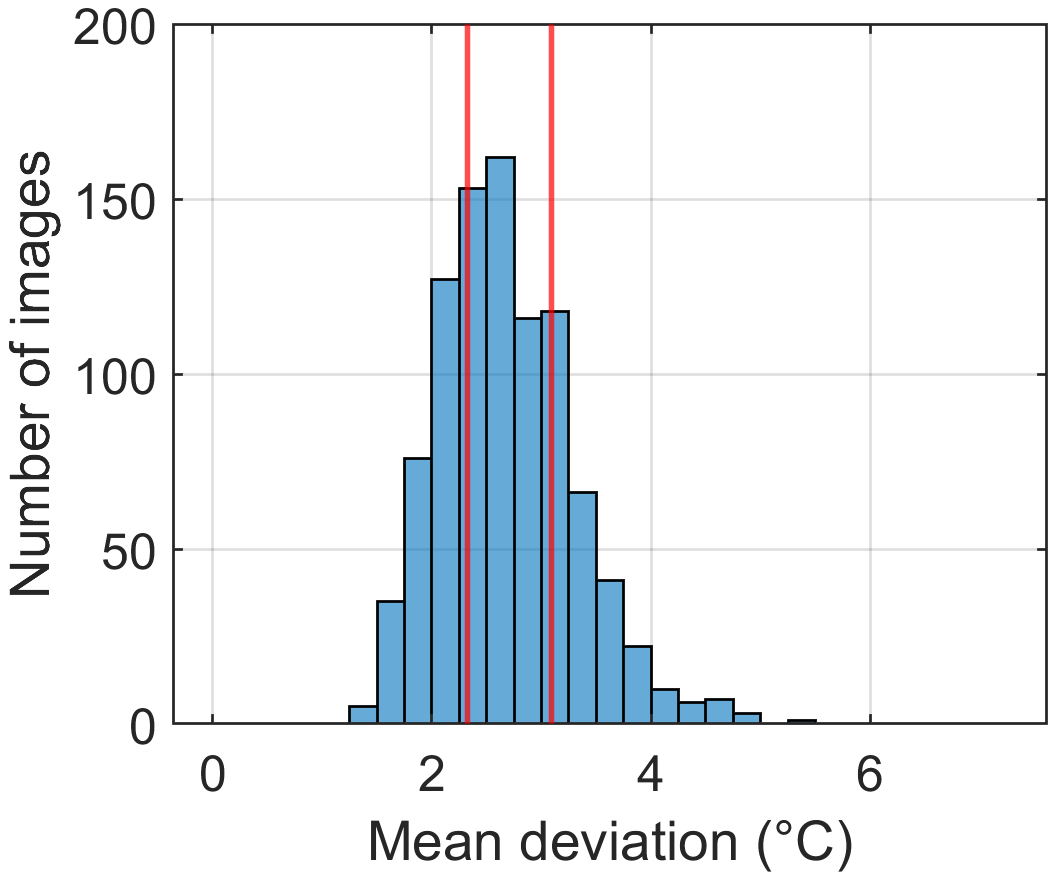}
        \caption{}
        \label{fig:stats_Sum_on_Wint4}
    \end{subfigure}
    \hfill
    \begin{subfigure}[b]{0.6\textwidth}
        \centering
        \includegraphics[width=\textwidth]{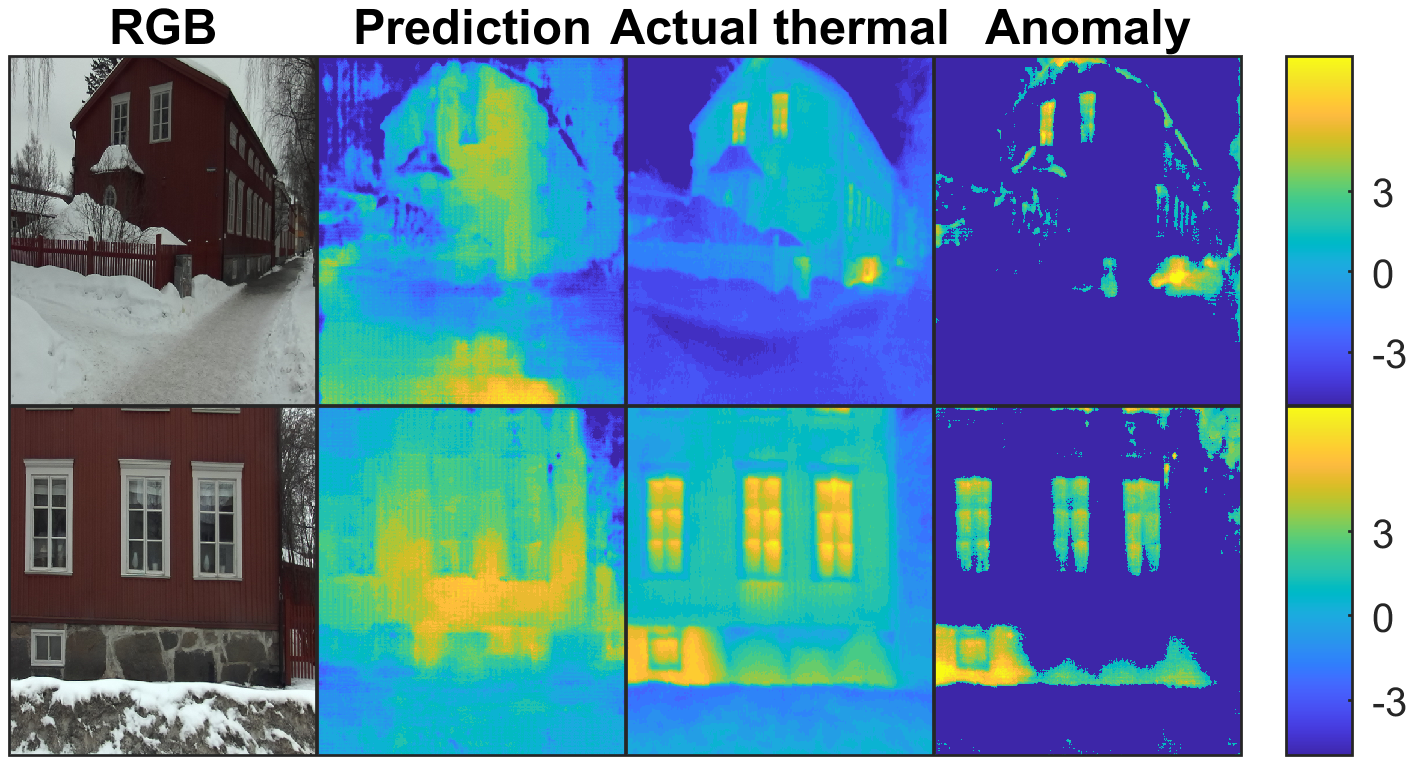}
        \caption{}
        \label{fig:prediction_Sum_on_Wint8}
    \end{subfigure}
    \hfill
    \begin{subfigure}[b]{0.35\textwidth}
        \centering
        \includegraphics[width=\textwidth]{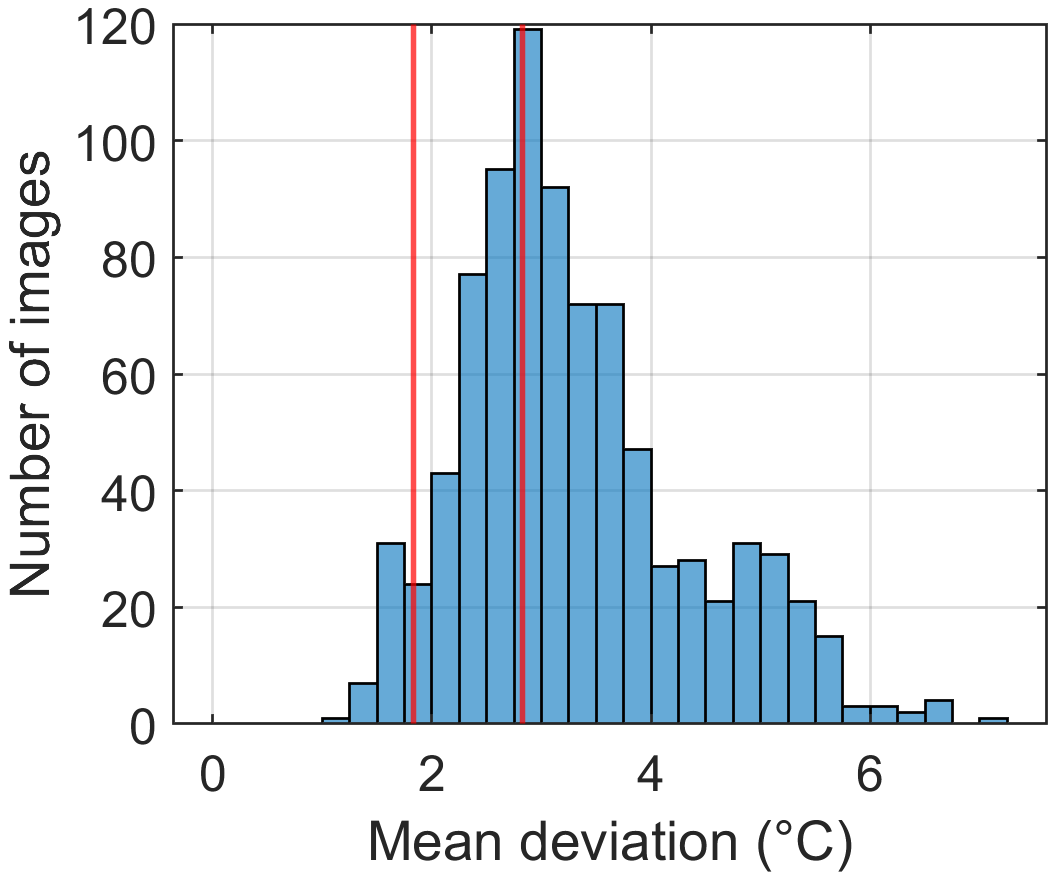}
        \caption{}
        \label{fig:stats_Sum_on_Wint8}
    \end{subfigure}
    \hfill
    \caption{Anomaly detection when using the (a,b) Winter4Net on Winter8 data, (c,d) SummerNet on winter4 data, and (e-f) SummerNet on winter4 data. The histogram plots show the average pixel deviation for all the images in the corresponding test sets, with the red vertical lines indicating the deviation for the displayed images. Scale bars in $^o$C over the outdoors temperature.}
    \label{fig:results_anomaly}
\end{figure}

To quantify the deviation between the predicted and actual thermal distributions, the mean of the absolute difference across the pixels was calculated for all the images in the test sets. The results are plotted as histograms in figures \ref{fig:stats_Net4_on_Net4}, \ref{fig:stats_Net8_on_Net8}, and \ref{fig:stats_sum_on_sum} with red vertical lines indicating the deviation of the displayed images for each case. The distribution peaks below $0.5$ degrees for the networks Winter4Net and Winter8Net, i.e., below the temperature resolution at a single pixel, but it seems to decay as a negative exponential, which might indicate potential poor predictions at its tail. Comparatively, the SummerNet shows a less accurate and reliable performance, although this was expected due to the significantly less available training data.

Next, we study how the network model performs when the environmental conditions for training and evaluation are different. This provides us with both insights into the generalization capabilities of the networks as well as the ability to detect anomalies that manifest in certain environmental conditions based on a network trained in alternative conditions. Figure \ref{fig:results_anomaly} shows the comparison between predicted and actual thermal distributions, as well as the thresholded difference between the two. Since we want the algorithm to recognize thermal bridges, $F$ in equation \ref{eq:difference} is taken as an identity, and in all shown cases a tolerance (T) of 1 $^o C$ to blank the difference and establish the anomaly was used. In figure \ref{fig:prediction_Wint4_on_Wint8}, the Winter4Net was used to evaluate the Winter8 data set. In this case, the difference between actual and predicted thermal images emphasized the windows of the facade and the snow, making the algorithm label them as anomalies. Quantitative assessments of windows based on thermal imaging is generally a challenging task due to the thermal radiation transmitted from the interior of the building, and thus the detected window anomaly here presented must be understood as a qualitative feature.

In figures \ref{fig:prediction_Sum_on_Wint4} and \ref{fig:prediction_Sum_on_Wint8}, the SummerNet is used to evaluate the Winter4 and Winter8 data sets, respectively. In this case, the model expects the ground and the walls to be hot, which discards them as anomaly as we defined it. However, most windows appeared hotter than predicted and labelled as anomalies. What is more, some basements presented areas with unusually high temperature in the actual thermal images, hinting at insulation faults a localized high indoors temperature, and were correctly labelled as anomalies.

\section{Conclusion}

We reported on a label-free method for the identification of anomalies in building envelopes. The results indicate that the pix2pix network model learned the correlation between color and thermal images and can make predictions of higher quality than were so far reported. We observed that an adequate data preprocessing before training was critical to obtain the desired results. When the algorithm is trained and evaluated in different conditions, it can identify relevant anomalies for the inspection of building envelopes. Specifically, we demonstrated this principle by recording training data at different outdoor temperatures, which thought the algorithm to detect thermal bridges. However, using data of targeted buildings can teach the algorithm to identify as normal the features of choice, such as those belonging to a specific energy efficiency classification.

With the proposed method, the gap between the controlled environment and the complexity of the outdoor scenario is minimized. Less labour intensive labeled data acquisition process facilitates smoother implementation in industrial applications. Furthermore the experimental protocol does not require capturing the wide range of variability of anomalies in the training data. This further makes the implementation of the suggested methods more realistic in industrial scenarios and enables further automation of thermal inspection.

\section*{Funding}
This work is supported by the Swedish Energy Agency (Energimyndigheten, grant No P2021-00202 and P2022-00141), and by the Swedish Research Council for Sustainable Development (Formas, grant No. 2022-01475).
\section*{Acknowledgments}
We thank Umeå Kommun, the Innovation Office (Innovationskontoret vid Umeå Universitet), and Uminova Innovation for their contributions and support.

\bibliographystyle{elsarticle-num}
\bibliography{references}

\end{document}